\documentclass[aps,pra,twocolumn,groupedaddress]{revtex4-1}
\usepackage{graphicx, listings}
\usepackage{epstopdf}
\usepackage{bbold}

\usepackage[latin1]{inputenc}
\usepackage{color}
\usepackage[a4paper,colorlinks=true,urlcolor=black,citecolor=black,linkcolor=black]{hyperref}
\usepackage{latexsym}
\usepackage{amsmath}
\usepackage{epstopdf}

\begin{document}

\title{Demixing Effects in Mixtures of Two Bosonic Species}
\author{F. Lingua}
\affiliation{Department of Applied Science and Technology and u.d.r. CNISM, Politecnico di Torino, I-10129 Torino, Italy}
\author{B. Capogrosso Sansone}
\affiliation{H. L. Dodge Department of Physics and Astronomy, The University of Oklahoma, Norman, Oklahoma, USA and\\
Department of Physics, Clark University, Worcester, Massachusetts 01610}
\author{M. Guglielmino}
\affiliation{Department of Applied Science and Technology and u.d.r. CNISM, Politecnico di Torino, I-10129 Torino, Italy}
\author{V. Penna}
\affiliation{Department of Applied Science and Technology and u.d.r. CNISM, Politecnico di Torino, I-10129 Torino, Italy}

\date{\today}

\begin{abstract}
Motivated by recent experiments on two-component systems, we investigate the ground-state
phase diagram of a mixture of two bosonic species by means of path-integral quantum Monte
Carlo by the two-worm algorithm. The mixture is trapped in a square lattice at different filling
conditions. Various quantum phases are stabilized depending on the interplay between intra-
and inter-species interactions and on the filling factors. We show that the ground-state phase diagram at half-filling features a demixed superfluid
phase and demixed Mott-Insulator phase when the inter-species interaction becomes
greater than the intra-species repulsion, and a double-superfluid phase or a supercounterflow
otherwise. We show that demixing, characterized by spatial separation of the two
species, can be detected experimentally through the effects of anisotropy revealed by
time-of-flight images. We also study how demixing effects depend on the filling factor of the two components. Finally, we found that super-counterflow phase is preserved in the presence of unbalanced populations.
\end{abstract}

\maketitle

\section{Introduction}
Mixtures of two bosonic species trapped in optical lattices
feature a variety of unprecedented
effects and quantum phases \cite{kuklSvitz}-\cite{demler} resulting from the interplay between kinetic energy and intra-, inter-species density-density interactions.
In the past decade, a considerable theoretical work has been devoted to investigating  manifold properties of these systems.
Different aspects of the phase diagram have been studied by means of generalized
mean-field schemes \cite{isacs}-\cite{pai12}, Luttinger-liquid picture \cite{mathey},
or perturbation methods \cite{iskin}. Moreover, the effect of phase separation \cite{mishra,boninsegni2011},
the study of quantum emulsions and coherence properties of mixtures \cite{rosci,buon},
and the shift of Mott domains due to the presence of a second (superfluid) species \cite{guglpenn_1} along with the interpretation of this shift in terms of polaron excitations \cite{demler} have been explored.

In the limit of large interactions, magnetic-like phases such as the incompressible
double-checkerboard solid and the supercounterflow have been predicted theoretically \cite{kuklSvitz}-\cite{altman} at total filling one and repulsive inter-species interaction, while paired superfluidity has been found \cite{KPS} in the case of attractive inter-species interaction at equal integer filling of the two components.
These findings stimulated further investigation of magnetic-like phases including finite temperature effects \cite{CapogrSoyler2010,Nakano_2012}, different optical lattice geometries~\cite{Ping_2014} and dimensionality, and various interacting
regimes \cite{argu}-\cite{Li}, \cite{iskin}.
Nevertheless, over ten years from initial theoretical investigation of these systems \cite{kuklSvitz}-\cite{altman},
their rich phase diagram
still exhibits several unexplored aspects that challenge theoretical and numerical techniques, while its elusive character demands more sophisticated experimental techniques for the observation of these quantum phases.

The recent experimental realization of mixtures, either combining two different atomic species
\cite{Catan_1,Thalm_1} or using the same atomic species in two different internal
energy states \cite{gadw_1,gadw_2,Sengstock1,Sengstock2} demonstrated how refined experimental techniques allow to control
the model parameters, hence reinforcing the interest toward these systems.
The possibility to observe such new phases in real systems is strongly
affected by {\it i}) the presence of the trapping potential which introduces an undesired source
of inhomogeneity, {\it ii}) the fact that their theoretic prediction is based on assuming
rather ideal conditions (such as, for example, species A and B with the same boson numbers $N_a = N_b$, or large intra-species interactions), and {\it iii}) the difficulty in
reaching low enough temperatures where such phases are expected.
Concerning points {\it i}) and {\it ii}), the experimental realizability of
magnetic phases has been analyzed in \cite{guglpenn_2} leading to promising results
at least for the double checkerboard phase. In Ref.~\cite{guglpenn_2} it was shown that the double checkerboard state can be found for
large but finite intra-species interaction and for certain parabolic confinements and particle number imbalance.

In this work, we reconstruct the ground-state phase diagram of a mixture of two
twin species (two bosonic components with equal hopping parameters and equal intra-species interactions) at half-filling $\nu=\frac{1}{2}$
and determine under which conditions the supercounterflow and demixed phases
are stabilized. We further explore demixing away from $\nu=\frac{1}{2}$. Demixed phases are characterized by spatial separation of the two species.
So far, demixing effects have been mainly studied in the context of continuous systems
\cite{D_param,segregPhase1998,Niklas_DemixingBEC2011,Niklas_DemixingBEC2014,
Saito2015,Saito2010,Pino2008} and Bose-Fermi mixtures \cite{Pollet1,Pollet2}.
Here, we study demixing in a binary mixture of bosons described by the two-component Bose-Hubbard model. At $\nu=\frac{1}{2}$ we find a demixed superfluid (dSF) phase or a demixed Mott-Insulator (dMI) phase
when the inter-species interaction is greater than the intra-species repulsion,
and a double-superfluid (2SF) phase or a supercounterflow (SCF) otherwise.
We characterize transitions to demixed phases by introducing a suitable demixing
parameter. Further significant information is
found by looking at the off-diagonal correlator which allows
to design experimental observation of the various phases through time-of-flight images. We also study SCF away from $\nu=\frac{1}{2}$ and find that SCF is stabilized for commensurate total filling although the superfluid response in the counter-flow channel does depend on the population imbalance.
%
This paper is organized as follows. In section \ref{sez1} we present our model and
formalism. In section \ref{res} we discuss the
ground state phase diagram at $\nu=\frac{1}{2}$, supercounterflow properties for unbalanced populations, and demixing effect at non-commensurate filling. Finally, in section \ref{concl} we conclude.

\section{Method and Model}\label{sez1}
We study a mixture of two bosonic species in a uniform two-dimensional (2D)
square optical lattice. The system is described by the two-component Bose-Hubbard (BH) model:
\begin{equation}
H = H_a + H_b + U_{ab} \sum_{i} n_{ai}n_{bi}
\label{H1}
\end{equation}
where $U_{ab}$ is the inter-species repulsion, $n_{ai,bi}$ is the number operator at site $i$ for species A and B, and
\begin{equation}
H_c = \frac{U_c}{2} \sum_{i}n _{ci}(n_{ci} - 1) - t_c\sum_{\langle ij\rangle} c_{i}^{\dag}c_{j}
\; ,
\label{Hc}
\end{equation}
with $c= a, b$ denoting the bosonic species, and operators
$c_i$, $c_i^\dag$ satisfying the standard commutator $[c_i, c_i^\dag]= 1$. Parameters $U_c$  represents the intra-species repulsion and $t_c$ is
the hopping amplitude
for component $c$. The symbol ${\langle ij\rangle}$ refers to
summation over nearest neighboring sites.
To further simplify the number of free parameters, we work with twin species, that is, we set $U_a = U_b= U$, $t_a = t_b= t$.
We are interested in exploring the phase diagram of model~(\ref{H1}) as a function of $U/t$ and $U_{ab}/t$ with particular emphasis on the supercounterflow and demixed phases. Our results are based on large-scale path-integral quantum Monte Carlo simulations by a two-worm algorithm~\cite{2worm}. Unless otherwise noted, we perform simulations for system sizes $L=8, 16, 24, 36$ (we use the lattice step $\lambda/2$ as a unit length) and we work at inverse temperature $\beta=L/t$ which ensures that the system is in its ground state.

\section{Results}\label{res}
\begin{figure}[h!]
\begin{center}
\includegraphics[trim={5cm 3cm 3cm 1cm}, clip, width=0.55\textwidth]{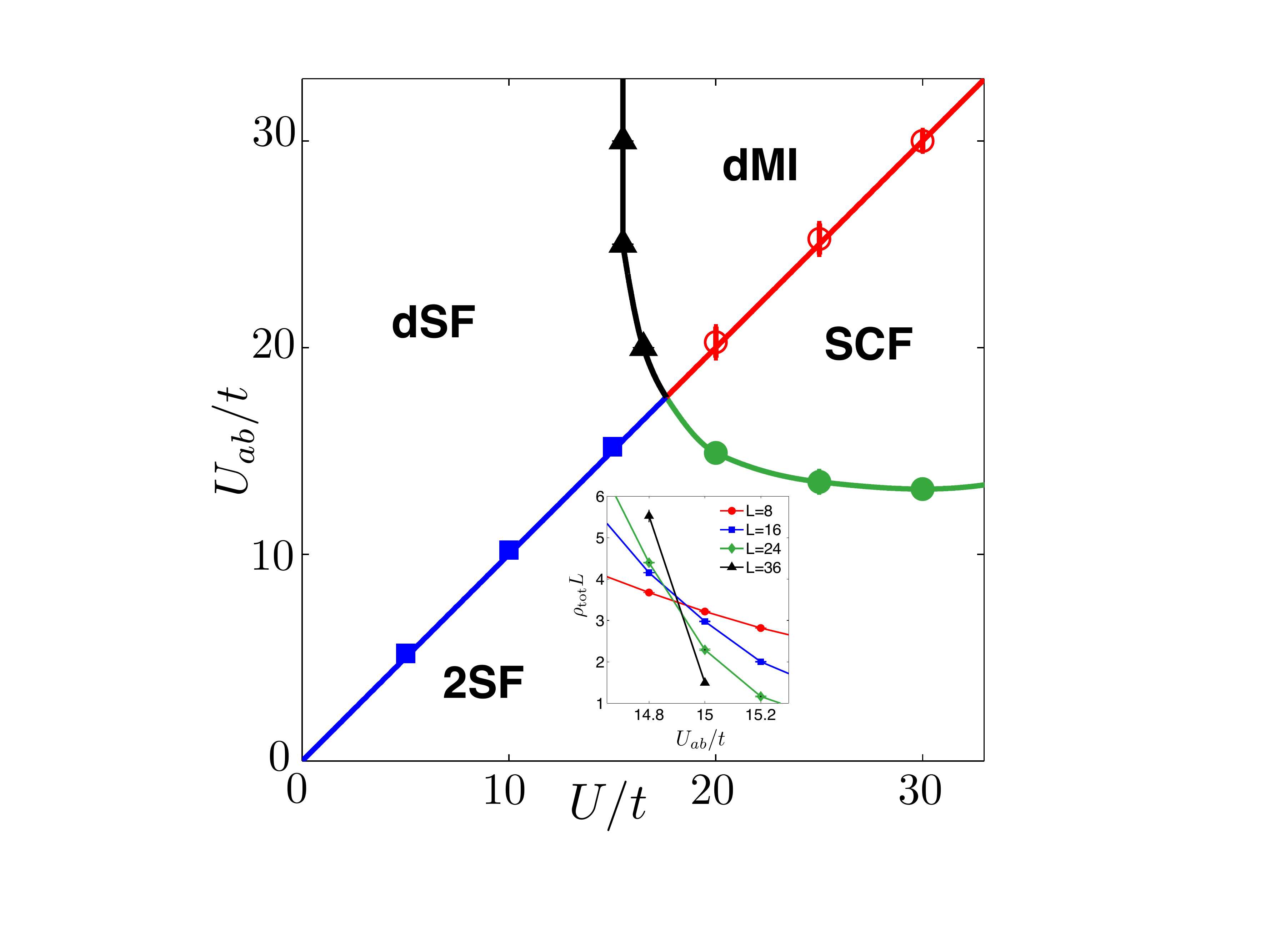}
\end{center}
 \caption{(Colors online) Ground-state phase diagram of twin bosonic species trapped in a uniform optical lattice at $\nu=\frac{1}{2}$. Four phases are stabilized: double superfluid (2SF), supercounterflow (SCF), demixed Mott-Insulator (dMI), and demixed superfluid (dSF). Symbols mark the phase boundaries as calculated from Monte Carlo simulations (see text). Whenever not visible, error bars lies within the symbol size. The inset shows the scaled total superfluidity $\rho_{\rm{tot}}$ as a function of the inter-species interaction for system sizes $L=8, 16, 24, 36$ and $U/t=20$. The intersection between curves corresponding to different system sizes gives the transition point for the 2SF-SCF transition. }\label{ph_diagr}
\end{figure}

{\em Ground-state phase diagram at half-filling}. The ground-state phase diagram of model (\ref{H1}) is shown in Fig. \ref{ph_diagr} in the $U/t$ vs $U_{ab}/t$ plane.
The 2SF phase features two $U(1)$ broken symmetries and is characterized by order parameters $\langle a\rangle\ne 0$ and $\langle b\rangle\ne 0$, or, equivalently, finite stiffness of the total superfluid flow $\rho_{\rm{tot}}\ne 0$, and finite stiffness of the relative superfluid flow $\rho_{\rm{SCF}}\ne 0$. The total and relative superfluid stiffnesses are given by $\rho_{\rm{tot,SCF}}={\langle (\vec W_a\pm\vec W_b)^2\rangle}/{2\beta}$, where $W_{a,b}$ are winding numbers of worldlines of species A and B~\cite{poll_cep_87}. The SCF phase restores one $U(1)$ broken symmetry and is characterized by order parameter $\langle a b^\dag\rangle\ne 0$ while $\langle a\rangle=0$ and $\langle b\rangle= 0$, or, equivalently, zero total superfluid stiffness, $\rho_{\rm{tot}}= 0$, and finite relative superfluid stiffness,  $\rho_{\rm{SCF}}\ne 0$. The demixed phases are characterized by spatial separation of the two components. This phenomenon is observed whenever $U_{ab}>U$ (as found in \cite{kuklSvitz} within the isospin picture of bosonic mixtures for the Mott region). A heuristic derivation of this condition for the case of generic filling factor is given in the Appendix A. In a demixed phase the density distribution of the two components is anisotropic on the lattice with density maxima of one species corresponding to density minima of the other.  The dSF
features two $U(1)$ broken symmetries, but the two species occupy different regions of the lattice. Finally, the dMI is an incompressible, insulating phase where the two components are spatially separated. Overall, 2SF and dSF are conducting phases, while dMI and SCF are insulating phases, although SCF supports flow in the so-called particle-hole channel. In agreement with \cite{comment2014}, we find that demixing is observed as soon as $U_{ab}/t>U/t$ for any value of $U$. It is also worth noting that in order to reach the insulating phases, both intra- and inter-species interactions have to be large enough. Indeed,  for $U_{ab}/t\lesssim 13.5$ ($U/t\lesssim 15.5$) the 2SF (dSF) phase is stable for arbitrarily large $U/t$ ($U_{ab}/t$).

In order to extract the phase diagram shown in Fig.~\ref{ph_diagr} we measure superfluid stiffness in terms of winding numbers statistics and the demixing parameter $\Delta$ (see below) which depends on the density distribution. Both observables are readily available within the path-integral formulation by the worm algorithm. The demixing parameter is given by:
\begin{equation}
\Delta=\frac{1}{M}\sum_i\Big[\frac{\langle n_{ai} \rangle - \langle n_{bi}\rangle}{\langle n_{ai}\rangle + \langle n_{bi}\rangle}\Big]^2 \label{delta}
\end{equation}
where $\langle n_{ci} \rangle$ is the quantum-thermal average of the density of component $c=a,b$ at site $i$. Parameter (\ref{delta}) is basically a lattice average of the square of the imbalance of the local species' density. This parameter is different than the one used in Ref.~\cite{D_param} $D=\Big[\frac{\langle N_a\rangle-\langle N_b\rangle}{\langle N_a\rangle+\langle N_b\rangle}\Big]^2$, but gives the same information about demixed phases.

In Fig.~\ref{ph_diagr}, solid circles correspond to the 2SF-SCF transition. This transition belong to the (2+1)-XY universality class. Transition points are found using standard finite-size scaling of $\rho_{\rm{tot}}$ as it can be seen in the inset of Fig.~\ref{ph_diagr} where we plot the scaled total superfluidity as a function of interaction $U_{ab}/t$ for system sizes $L=8, 16, 24, 36$ and $U/t=20$. The curves corresponding to different sizes intersect at the critical point $U_{ab}/t$ $=14.9\pm0.1$. Both 2SF and SCF are stable for $U_{ab}<U$.

\begin{figure}[h]
\begin{center}
\includegraphics[width=\columnwidth]{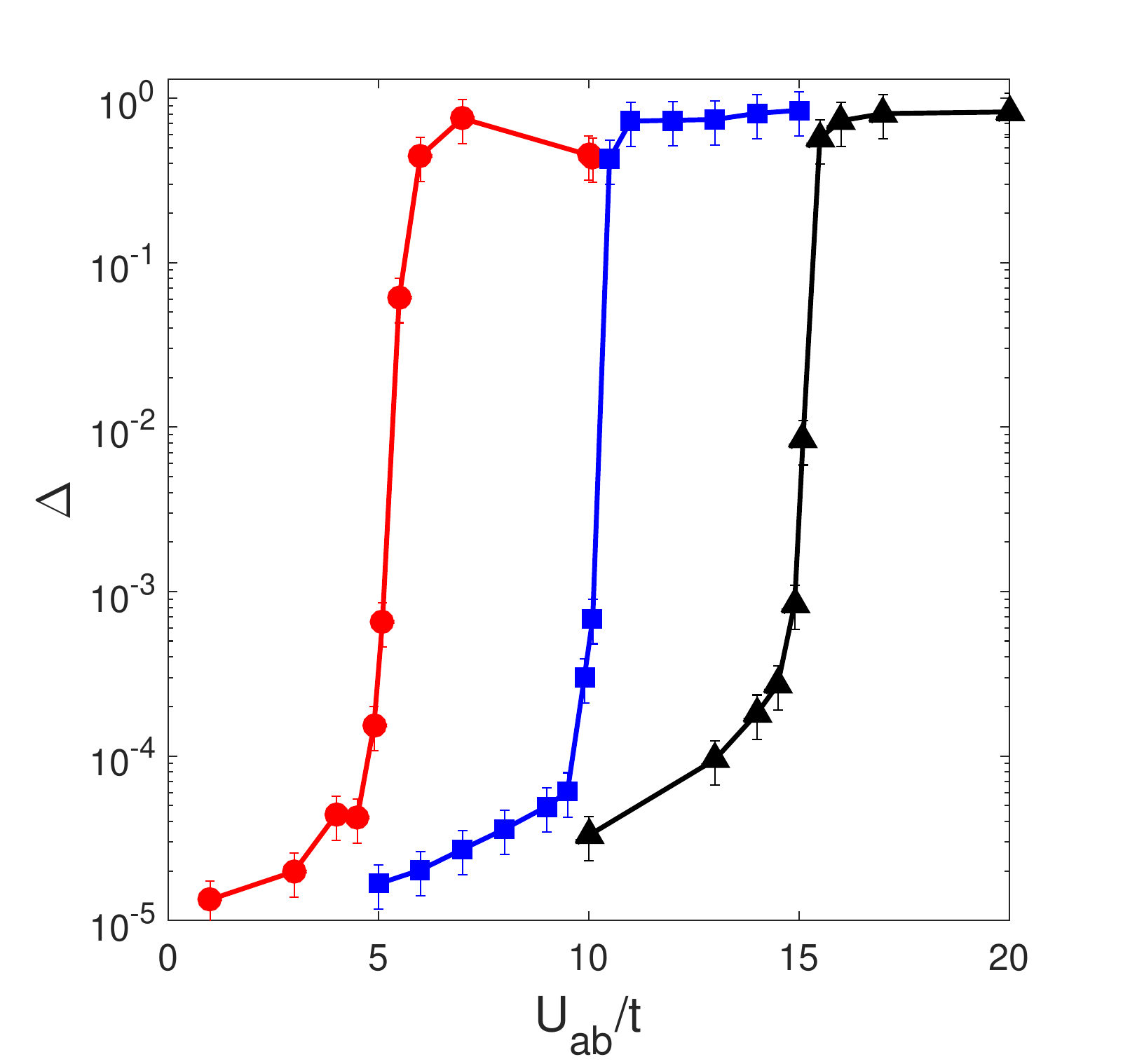}
\end{center}
 \caption{(Colors online) Demixing parameter $\Delta$ as function of $U_{ab}/t$ for $U/t=5, 10, 15$ (circles, squares, triangles respectively) across the 2SF-dSF transition. A jump of four orders of magnitude is clearly visible when $U_{ab}/t\sim U/t$ signaling the onset of the demixed phase dSF at the expense of the 2SF.}\label{del_2SF2dSF}
\end{figure}

\begin{figure}[h!]
\includegraphics[trim={2cm 1cm 3cm 1cm}, clip, width=0.5\textwidth]{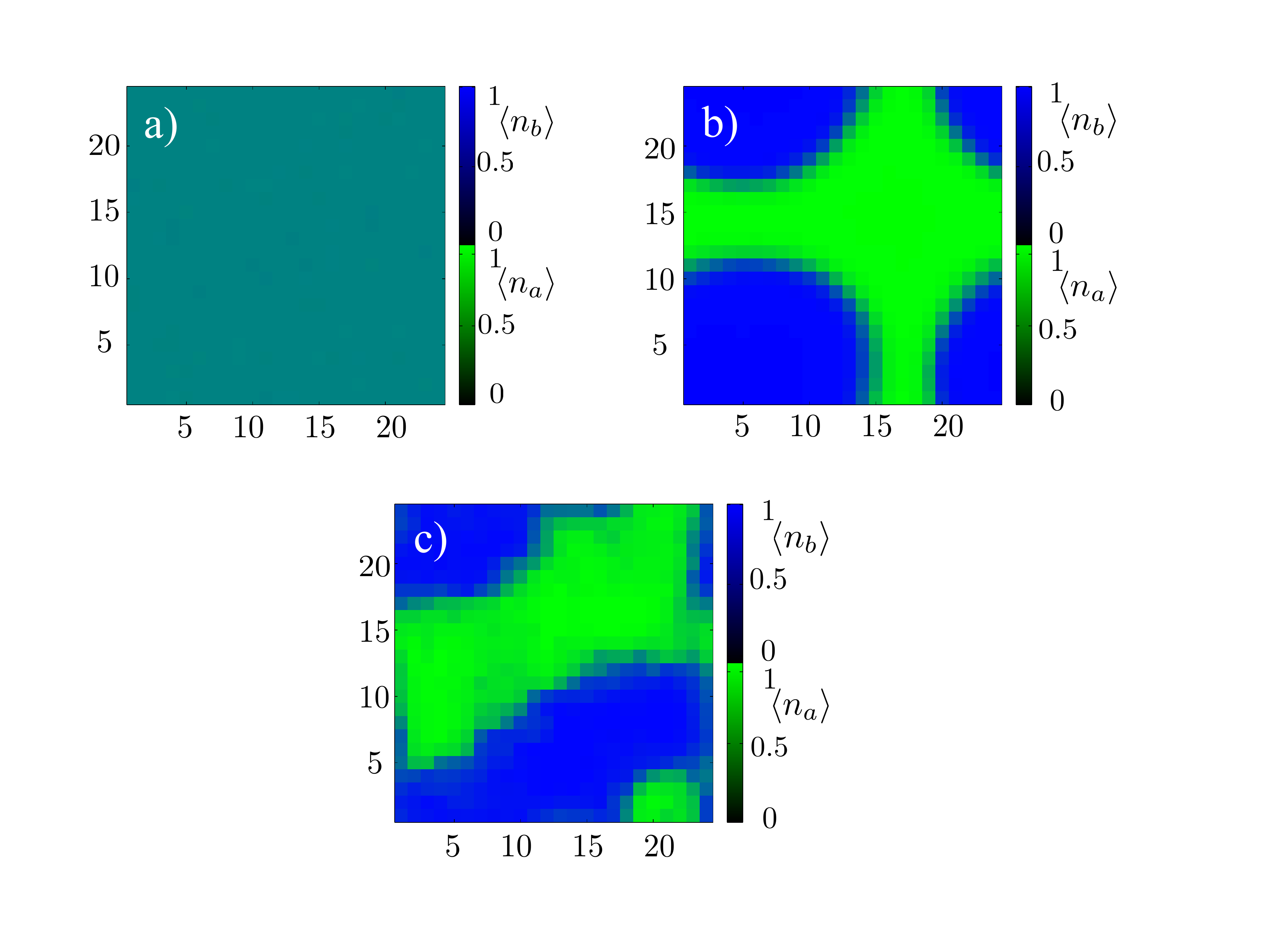}
\caption{(Colors online) Quantum-statistical average of the particle number $\langle n_{ai}\rangle$, $\langle n_{bi}\rangle$. The $x$ and $y$ axis denote the $x$ and $y$ coordinates on the lattice. The color code is displayed on the the right bar. Panel a): 2SF phase at $U/t=10$, $U_{ab}/t=7$ with density uniformly distributed in the lattice. Panel b): dSF phase at $U/t=15$, $U_{ab}/t=20$ with two components occupying spatially-separated regions of the lattice. Panel c): dMI phase at $U/t=20$, $U_{ab}/t=22$ with two components occupying spatially-separated regions of the lattice. Compenetration between the two regions is noticeable.}\label{ph_sep}
\end{figure}

We detect the phase transition between 2SF and dSF (squares in Fig.~\ref{ph_diagr}) by studying the behavior of the $\Delta$ parameter. In Fig. \ref{del_2SF2dSF} we show $\Delta$ as a function of $U_{ab}/t$ for $U/t=5$, $U/t=10$, $U/t=15$, circles, squares, triangles respectively. A jump of four orders of magnitude is clearly visible when $U_{ab}/t\sim U/t$ signaling the onset of the demixed phase dSF at the expense of the 2SF. Further significant information about the demixed phase is achieved by looking at the density distribution of particles in the lattice. The quantum-statistical average of the particle number $\langle n_{ai}\rangle$, $\langle n_{bi}\rangle$ is displayed in Fig.~\ref{ph_sep}. The $x$ and $y$ axis denote the $x$ and $y$ coordinates on the lattice. The color code is displayed on the the right bar. Panel a) refers to the 2SF case where the density of each species is uniformly distributed in the lattice, corresponding to the spatial coexistence of the two species. Panel b) refers to the dSF phase. Here, we clearly see that the two components occupy spatially-separated regions with well defined boundaries of a few lattice steps of thickness where the two components coexist.

The dSF phase becomes unstable towards a dMI phase upon increasing the intra-species interaction (triangles in Fig.~\ref{ph_diagr}). We compute the phase boundary by verifying the drop of the SF density of each component for system sizes $L=8, 16, 24$. While the dSF is characterized by two U(1) broken symmetries over spatially separated regions, in the dMI these symmetries are restored, that is, the system loses its off-diagonal long range (anisotropic) correlations and becomes insulating. Investigating the details of this phase transition is challenging. Finite-size scaling of the SF stiffness cannot be performed in proximity of
demixed regions since the statistics of winding numbers is affected by the topography and the (non-connected vs. connected) topology of demixed regions, both of which depend on the initial conditions. The averaged particle number within the lattice in the dMI phase is displayed in panel c) of Fig.~\ref{ph_sep}. Unlike the dSF phase (panel b)), the boundaries between the regions occupied by the two species tend to be rough and irregular, and compenetration between the regions is more pronounced due to the reduced mobility of bosons in the dMI phase.

Finally the empty circles in Fig.~\ref{ph_diagr} correspond to the SCF-dMI transition. Upon entering the dMI the system restores the U(1) broken symmetry characterizing the SCF. We measured the $\Delta$ parameter and $\rho_{\rm{SCF}}$ across the transition line and, similarly to the 2SF-dSF transition, we observe an increase of about three orders of magnitude in $\Delta$, while $\rho_{\rm{SCF}}$ goes to zero.

\begin{figure}[h!]
\includegraphics[trim={2cm 1cm 3cm 1cm}, clip, width=0.5\textwidth]{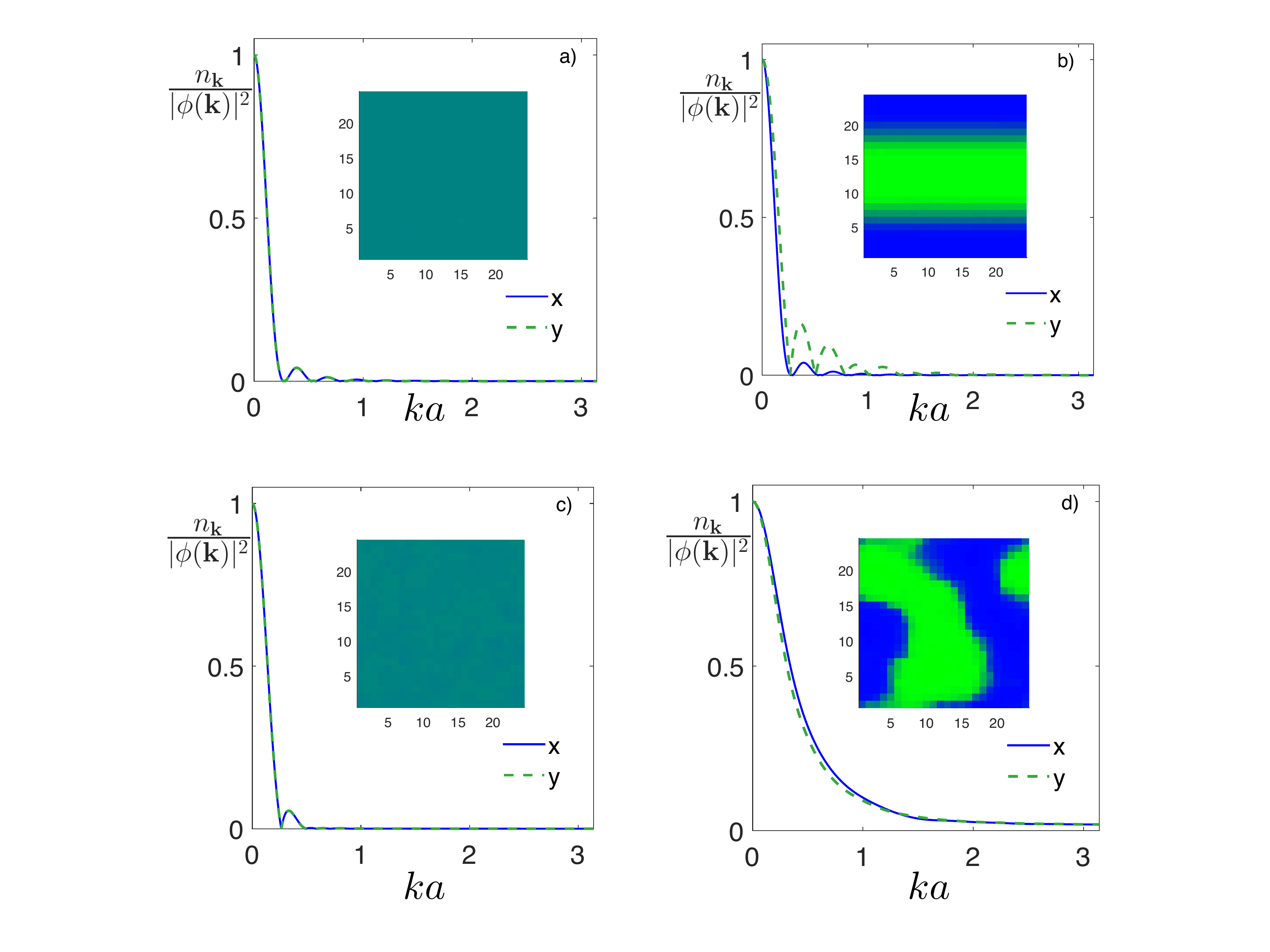}\caption{Momentum distributions $n_k$ in the first Brillouin zone for double superfluid 2SF (panel a)), demixed superfluid dSF (panel b)), supercounterflow (panel c)) and demixed Mott-Insulator (panel d)). The insets show the corresponding quantum-statistical average of the density of the two components within the lattice. The demixed phases dSF shows an evident anisotropy along the $x-$ and $y-$direction. This is expected due to the anisotropy of the spatial separation. }\label{fig4}
\end{figure}

In figure \ref{fig4} we show the computed momentum distributions $n_{c \mathbf{k}}=|\phi_{(\mathbf{k})}|^2 \sum_{i,j}e^{i\mathbf{k}(\mathbf{r}_i - \mathbf{r}_j)}\langle c^{\dag}_i c_j\rangle $ \cite{tof1} for species $c=a,b$ proportional to Time-Of-Flight (TOF) images detectable experimentally. Here $\phi_{(\mathbf{k})}$ is the Fourier transform of Wannier function $\phi(\mathbf{r})$, which we do not compute here. TOF profiles along $x$ and $y$ lattice directions within the first Brillouin zone are plotted in figure~\ref{fig4} for the 2SF (panel a)), dSF (panel b)), SCF(panel c)) and dMI (panel d)). Note that the TOF image of SCF corresponds to the one of the particle-hole pair. The insets show the corresponding quantum-statistical average of the density of the two components within the lattice. We observe different profiles featuring the different quantum phases with an evident detectable anisotropy of distributions $n_{c \mathbf{k}}$ along the $x-$ and $y-$direction for dSF. This is expected due to the anisotropy of the spatial separation and represents a clear experimental signature of the dSF. On the other hand the TOF image of dMI doesn't reflect anisotropy of spatial separation due to the absence of off-diagonal long range order.
%
%
In general, we observe no differences in the momentum distributions between the two species.

\begin{figure}[h!]
\includegraphics[width=0.5\textwidth]{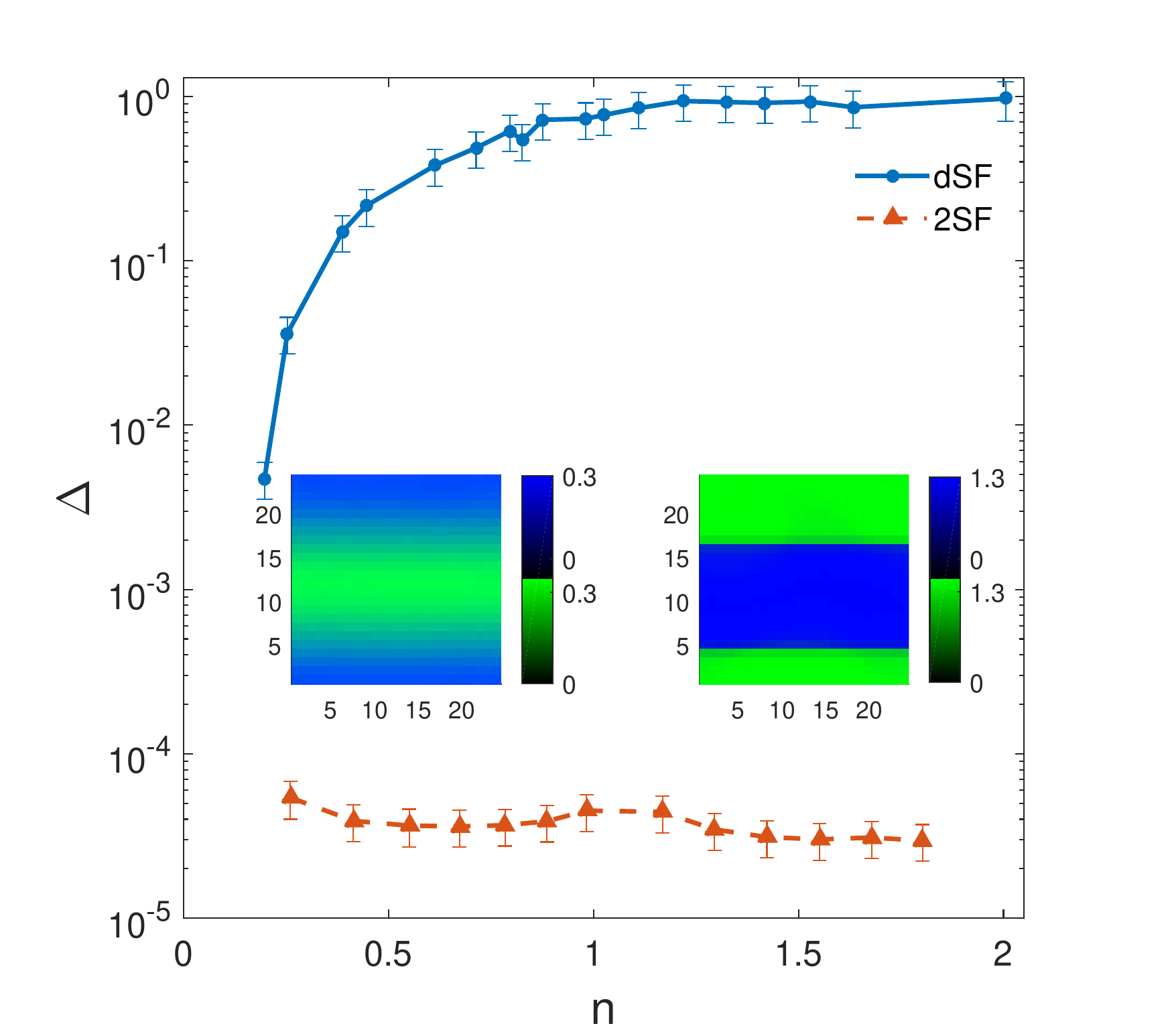}\caption{$\Delta$ parameter as a function of total filling $n$ for dSF (circles, $U/t=10$, $U_{ab}/t=15$ and $L=24$) and 2SF (triangles, $U/t=10$, $U_{ab}/t=7$ and $L=24$) phases. Insets represent the average densities of the two components for $n=0.3$ (left), $n=1.3$ (right).} \label{fig5}
\end{figure}

{\em Non-Commensurate Filling}. We now turn to studying model~(\ref{H1}) away from commensurate total filling where neither the dMI nor the SCF exist. In particular, we are interested in studying the properties of the dSF phase as a function of the total filling $n$. We find that demixing effects are still present at $n<1$ although the spatial separation between the two components A, B is not as pronounced as for $n\ge1$. This can be seen in Fig.~\ref{fig5}, where we plot the demixing parameter $\Delta$ as a function of $n$ (circles). A substantial drop in the value of $\Delta$ is observed for $n<1$. This is due to the presence of large regions in the lattice where A and B overlap, as shown in the left inset of  Fig.~\ref{fig5} where we plot the quantum statistical average of the densities of the two species at $n=0.3$. This overlap region results from enhanced hopping of particles which is responsible for larger fluctuations of $\langle n_{a,b}\rangle$. For comparison, in the right inset of Fig.~~\ref{fig5}, we show the species densities at $n=1.3$. A net spacial separation between A and B is observed. Despite this substantial drop in the value of $\Delta$ for $n<1$, we find that $\Delta$ is still a good indicator that demixing has occurred. Indeed, $\Delta$ values in the 2SF phase (triangles in Fig.~\ref{fig5}) are still orders of magnitude smaller than in the dSF phase. These results suggests that demixing effects can be observed in the presence of an external harmonic trap where a variation of $n$ within the trap is present. See Supplemental Material for further details.

\begin{figure}[h!]
\includegraphics[width=0.5\textwidth]{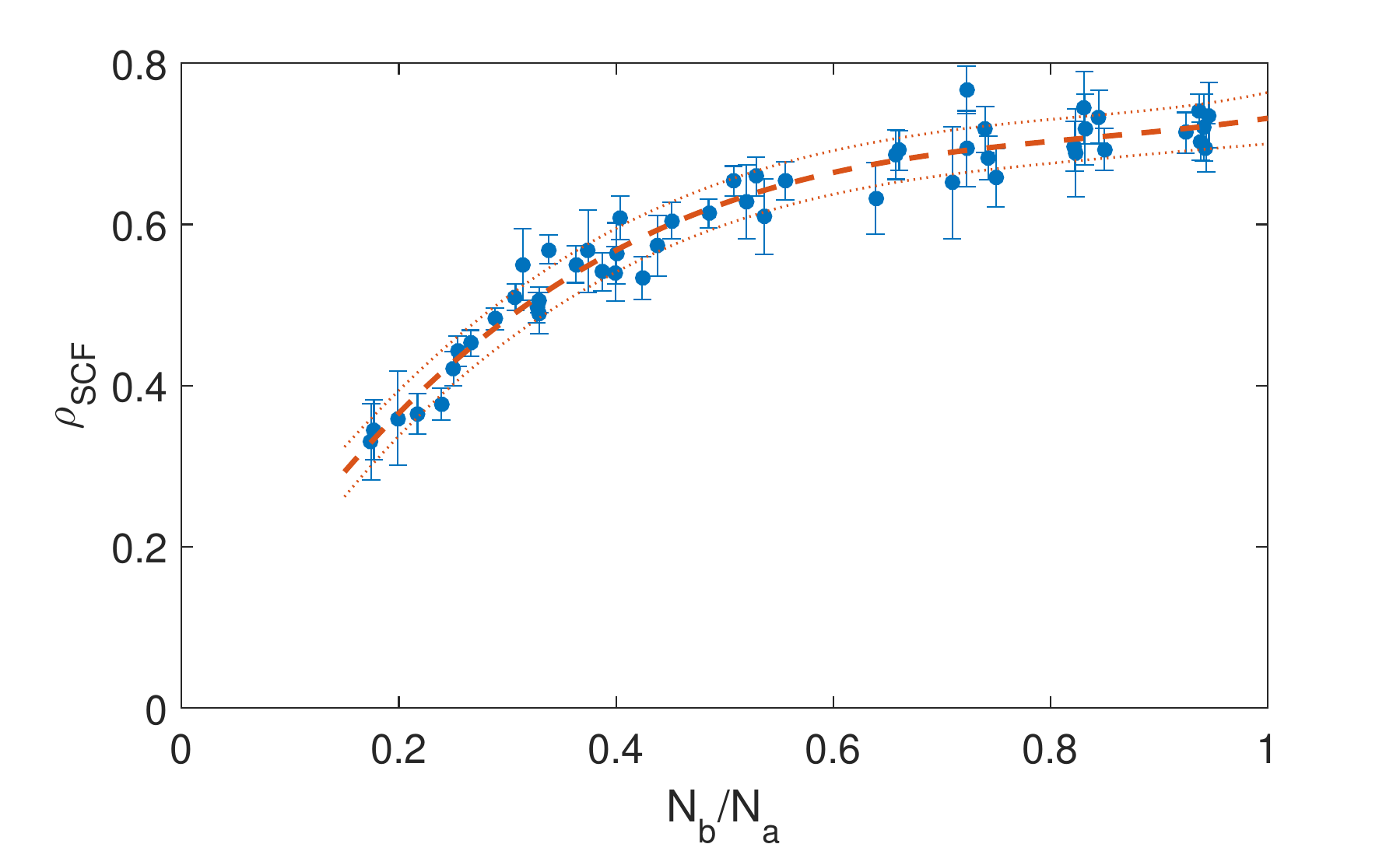}\caption{Super-counterfluidity stiffness as a function of $N_b/N_a$ ($U/t=20$, $U_{ab}/t=17$ and $L=24$). Dashed line is a cubic polynomial least-square fit, while dotted lines represent one-$\sigma$ deviation contours from the fit.}\label{fig6}
\end{figure}

{\em Unbalanced Populations}. We conclude by studying the SCF in the presence of population imbalance. The SCF phase can be stabilized at $n=1$. Here we are interested in showing that SCF still exists with non-zero imbalance although its robustness depends on the latter. In Fig. \ref{fig6} we plot $\rho_{SCF}$ for different values of ratio $N_b/N_a$. We observe that SCF remains robust also for large population imbalance although the largest superfluid response corresponds to the balanced case.

\section{Conclusion}\label{concl}
Motivated by recent experiments on two-component systems, we have investigated properties of a mixture trapped in a 2D optical lattice at different filling factors. We have used path-integral quantum Monte Carlo by the two-worm algorithm. At $\nu=\frac{1}{2}$, we have observed a demixed superfluid phase and demixed Mott-Insulator phase when the inter-species interaction becomes greater than the intra- species repulsion, and a double superfluid phase or a supercounterflow otherwise. Demixing is characterized by spatial separation of the two components and manifests itself experimentally with anisotropic time of flight images. The latter have been calculated for selected examples, comparing the results for demixed and non-demixed phases. Finally, we have shown that the SCF phase survives in the presence of population imbalance and trapping potential (see Supplemental Material). In the future, we plan to investigate how finite temperature affects demixing. Our preliminary results show that, in the dSF phase, thermal fluctuations destroy demixing effects earlier than they destroy superfluidity, leaving the system in a uniform superfluid phase. Moreover, since preliminary results show that the demixing parameter is sensitive to temperature, this suggests its possible application as a thermometer for mixtures of ultracold gases.

{\em Acknowledgements} We would like to thank F. Minardi for fruitful discussions. This work was supported by MIUR (PRIN 2010LLKJBX) and by the NSF (PIF-1415561). The computing for this project was performed at the OU Supercomputing Center for Education and Research (OSCER) at the University of Oklahoma (OU).


\begin{appendix}
\section{A simple interpretation of demixing effect}

The mixing/demixing effect can be interpreted in a simple way for the dSF-SF transition at generic filling factor.
In a perturbative framework, where the energy contribution of tunneling processes is assumed to be negligible,
one should compare the energy of the ground-state with the two species spatially separated to the energy of the ground-state where the two species coexist within the lattice. When the two species occupy spatially separated regions of the lattice
${\cal R}_a$ and ${\cal R}_b$, we assume that $M_a$ ($M_b$)
sites of ${\cal R}_a$ (${\cal R}_b$) are occupied by bosons A (B),
with $r_a$ ($r_b$) sites containing $n+1$ ($m+1$) bosons, and $M_a-r_a$ ($M_b-r_b$) sites
containing $n$ ($m$) bosons. Note that the total number of sites is given by
$M = M_a+M_b$ while the total number of particles is $N_a = M_a n + r_a $, $N_b = M_b m + r_b $. The nonuniform filling in
${\cal R}_a$ and ${\cal R}_b$ reflects the SF character of the two species.
The resulting energy reads:
\begin{eqnarray}
&E_0& = \frac{U_a}{2} M_a n(n-1) + \frac{U_b}{2} M_b m(m-1)
\nonumber
\\
&+&U_a r_a n + U_b r_b m -\mu_a N_a - \mu_b N_b\, ,
\end{eqnarray}
where the $U_{ab}$-dependent term is absent due to the spatial separation of the two species. The mixing effect is described by: a boson is lost from each
site of ${\cal R}_a$ (${\cal R}_b$) occupied by $n+1$ ($m+1$) bosons while $r_a$ ($r_b$) sites appear in ${\cal R}_b$
(${\cal R}_a$) with $m$ bosons B and one boson A ($n$ bosons A and one boson B). $U_{ab}$ interaction term is now activated and the resulting energy is
\begin{eqnarray}
&E_0'& = \frac{U_a}{2} M_a n(n-1) + \frac{U_b}{2} M_b m(m-1)
\nonumber
\\
&+& U_{ab}( r_a m + r_b n)  -\mu_a N_a - \mu_b N_b\, .
\end{eqnarray}
This mutual exchange of bosons between ${\cal R}_a$ and ${\cal R}_b$
represents the mixing process with the lowest-energy cost in the minimum-energy scenario.
The condition $E_0 < E_0'$ (justifying the transition from the uniform ground state to the demixed state)
implies that $U_{ab}( r_a m + r_b n)>U_a r_a n + U_b r_b m $ which reduces to the well-known condition
$U_{ab} > U$ for $n=m$ and $U_a=U_b$. This elementary argument is valid in the SF regime due to
its semiclassical character. It cannot be extended to the transition
from the dMI-SCF phase where quantum correlations and hopping processes play a
preminent role.

\end{appendix}


\begin{thebibliography}{}



\bibitem{kuklSvitz}
A. B. Kuklov and B.V. Svistunov, Phys. Rev. Lett. 90, 100401 (2003).

\bibitem{DDL}
L.-M. Duan, E. Demler and M. D. Lukin, Phys. Rev. Lett. 91, 090402 (2003).

\bibitem{altman}
E. Altman, W. Hofstetter, E. Demler, and M. D. Lukin,
New J. Phys. 5, 113 (2003).

\bibitem{KPS}
A. B. Kuklov, N. Prokof'ev, and B. V. Svistunov, Phys. Rev. Lett. 92, 050402 (2004).

\bibitem{argu}
A. Arg\"uelles and L. Santos, Phys. Rev. A 75, 053613 (2007).

\bibitem{hu}
A. Hu, L. Mathey, I. Danshita, E. Tiesinga, C. J. Williams, and C. W. Clark,
Phys. Rev. A 80, 023619 (2009).

\bibitem{hubener}
A. Hubener, M. Snoek, and W. Hofstetter, Phys. Rev. B 80, 245109 (2009).

\bibitem{2worm}
S. G. S\"oyler, B. Capogrosso-Sansone, N. V. Prokof'ev, B. V. Svistunov, New J. Phys. 11, 073036 (2009).
%

\bibitem{CapogrSoyler2010}
B. Capogrosso-Sansone, S. G. S\"oyler, N. V. Prokof'ev, and B. V. Svistunov, Phys. Rev. A 81, 053622 (2010).

\bibitem{oghoe} T. Oghoe, and N. Kawashima, Phys. Rev. A 83, 023622 (2011).

\bibitem{chung}
C.-M. Chung, S. Fang, and P. Chen, Phys. Rev. B 85, 214513 (2012).

\bibitem{Li}
Y. Li, L. He, and W. Hofstetter, New J. Phys. 15, 093028 (2013).

\bibitem{Nakano_2012}
Y. Nakano, T. Ishima, N. Kobayashi, T. Yamamoto, I. Ichinose and T. Matsui,
Phys. Rev. A 85, 023617 (2012).

\bibitem{Ping_2014}
J.-P. Lv, Q.-H. Chen, and Y. Deng, Phys. Rev. A 89, 013628 (2014).


\bibitem{isacs}
A. Isacsson, Min-Chul Cha, K. Sengupta, and S. M. Girvin, Phys. Rev. B 72, 184507 (2005).

\bibitem{guglpenn_3}
M. Guglielmino, V. Penna and B. Capogrosso-Sansone, Laser Physics 21, 1443 (2011)

\bibitem{pai12}
R. V. Pai, J. M. Kurdestany, K. Sheshadri, R. Pandit, Phys. Rev. B 85, 214524 (2012).

\bibitem{mathey}
L. Mathey, Phys. Rev. B 75, 144510 (2007).

\bibitem{iskin}
M. Iskin, Phys. Rev. A 82, 033630 (2010).


\bibitem{mishra}
T. Mishra, R. V. Pai, and B. P. Das, Phys. Rev. A 76, 013604 (2007).

\bibitem{boninsegni2011}
P. Jain and M. Boninsegni Phys. Rev. A 83, 023602 (2011)

\bibitem{rosci}
T. Roscilde and J. I. Cirac, Phys. Rev. Lett. 98, 190402 (2007).

\bibitem{buon}
P. Buonsante, S. M. Giampaolo, F. Illuminati, V. Penna, and A. Vezzani,
Phys. Rev. Lett. 100, 240402 (2008).

\bibitem{guglpenn_1}
M. Guglielmino, V. Penna and B. Capogrosso-Sansone, Phys. Rev. A 82, 021601(R) (2010).

\bibitem{demler}
D. Benjamin and E. Demler, Phys. Rev. A 89, 033615 (2014).

\bibitem{Catan_1}
J. Catani, L. De Sarlo, G. Barontini, F. Minardi, and M. Inguscio,
Phys. Rev. A 77, 011603(R) (2008).

\bibitem{Thalm_1}
G. Thalhammer, G. Barontini, L. De Sarlo, J. Catani, F.Minardi and M. Inguscio,
Phys. Rev. Lett. 100, 210402 (2008).

\bibitem{gadw_1}
B. Gadway, D. Pertot, R. Reimann, and D. Schneble, Phys. Rev. Lett. 105, 045303 (2010).

\bibitem{gadw_2}
B. Gadway, D. Pertot, J. Reeves, D. Schneble, Nature Physics 8, 544 (2012)

\bibitem{guglpenn_2}
M. Guglielmino, V. Penna and B. Capogrosso-Sansone, Phys. Rev. A 84, 031603(R) (2011).
%
%
%

\bibitem{D_param}
P. Jain, S. Moroni, M. Boninsegni, L. Pollet, Phys. Rev. A 88, 033628 (2013).

\bibitem{Bakr_2009}
W. S. Bakr, J. I. Gillen, A. Peng, S. Folling and M. Greiner, Nature 462, 74 (2009).

\bibitem{Bloch2010}
J. F. Sherson, C. Weitenberg, M. Endres, M. Cheneau, I. Bloch and S.Kuhr, Nature 467, 69 (2010)

\bibitem{bakr2015}
L. W. Cheuk, M. A. Nichols, M. Okan, T. Gersdorf, V. V. Ramasesh, W. S. Bakr, T. Lompe and M. W. Zwierlein, Phys. Rev. Lett. 114, 193001 (2015)

\bibitem{greiner2015}
M. F. Parsons, F. Huber, A. Mazurenko, C. S. Chiu, W. Setiawan, K. Wooley-Brown, S. Blatt, and M. Greiner, Phys. Rev. Lett. 114, 213002 (2015)

\bibitem{kuhr2015}
E. Haller, J. Hudson, A. Kelly, D. A. Cotta, B. Peaudecerf, G. D. Bruce and S. Kuhr, arXiv:1503.02005v2

\bibitem{Sengstock1}
P. Soltan-Panahi, J. Struck, P. Hauke, A. Bick, W. Plenkers, G. Meineke, C. Becker,	P. Windpassinger, M. Lewenstein	and K. Sengstock, Nature Physics 7, 434 (2011)

\bibitem{Sengstock2}
P. Soltan-Panahi, D. S. Luhmann, J. Struck, P. Windpassinger and K. Sengstock, Nature Physics 8, 71 (2012)	

\bibitem{Saito2015}
Y. Eto, M. Kunimi, H. Tokita, H. Saito, and T. Hirano, arXiv:1505.04882v1

\bibitem{Saito2010}
S. Tojo, Y. Taguchi, Y. Masuyama, T. Hayashi, H. Saito and T. Hirano, Phys. Rev. A 82, 033609 (2010).

\bibitem{Pino2008}
S. B. Papp, J. M. Pino, and C. E. Wieman, Phys. Rev. Lett. 101, 040402 (2008).

\bibitem{Niklas_DemixingBEC2011}
E. Nicklas, H. Strobel, T. Zibold, C. Gross,y B. A. Malomed, P. G. Kevrekidis, and M. K. Oberthaler, Phys. Rev. Lett. 107, 193001 (2011)

\bibitem{Niklas_DemixingBEC2014}
E. Nicklas, W. Muessel, H. Strobel, P. G. Kevrekidis, M. K. Oberthaler, arXiv: 1407.8049v1


\bibitem{comment2014}
F. Zhan and I. P. McCulloch, Phys. Rev. A 89, 057601 (2014)

\bibitem{segregPhase1998}
P. Ao and S. T. Chui, Phys. Rev. A 58, 6 4836 (1998)

\bibitem{Pollet1}
L. Pollet, M. Troyer, K. Van Houcke and S. M. A. Rombouts, Phys. Rev. Lett. 96, 190402 (2006)

\bibitem{Pollet2}
L. Pollet, C. Kollath, U. Schollwock and M. Troyer, Phys. Rev. A 77, 023608 (2008)

\bibitem{prkof_1}
N. V. Prokof'ev, B. V. Svistunov, I. S. Tupitsyn, Phys. Lett. A 238, 253 (1998).

\bibitem{prkof_2}
N. V. Prokof'ev, B. V. Svistunov, I. S. Tupitsyn, Zh. Eksp. Teor. Fiz. 114, 570 (1998).
\bibitem{poll_cep_87} E. L. Pollock and D. M. Ceperley, Phys. Rev. B 36, 8343 (1987).

\bibitem{tof1}
V. A. Kashurnikov, N. V. Prokof'ev and B. V. Svistunov Phys. Rev. A 66, 031601 (2002)

\end{thebibliography}
\end{document}


\title{Supplemental Material: Demixing Effects in Mixtures of Two Bosonic Species}
\author{F. Lingua}
\affiliation{Department of Applied Science and Technology and u.d.r. CNISM, Politecnico di Torino, I-10129 Torino, Italy}
\author{B. Capogrosso Sansone}
\affiliation{H. L. Dodge Department of Physics and Astronomy, The University of Oklahoma, Norman, Oklahoma, USA and\\
Department of Physics, Clark University, Worcester, Massachusetts 01610}
\author{M. Guglielmino}
\affiliation{Department of Applied Science and Technology and u.d.r. CNISM, Politecnico di Torino, I-10129 Torino, Italy}
\author{V. Penna}
\affiliation{Department of Applied Science and Technology and u.d.r. CNISM, Politecnico di Torino, I-10129 Torino, Italy}

\date{\today}

\pacs{}

\maketitle

\begin{figure}[h!]
\begin{minipage}[t]{.49\textwidth}
\includegraphics[width=\textwidth]{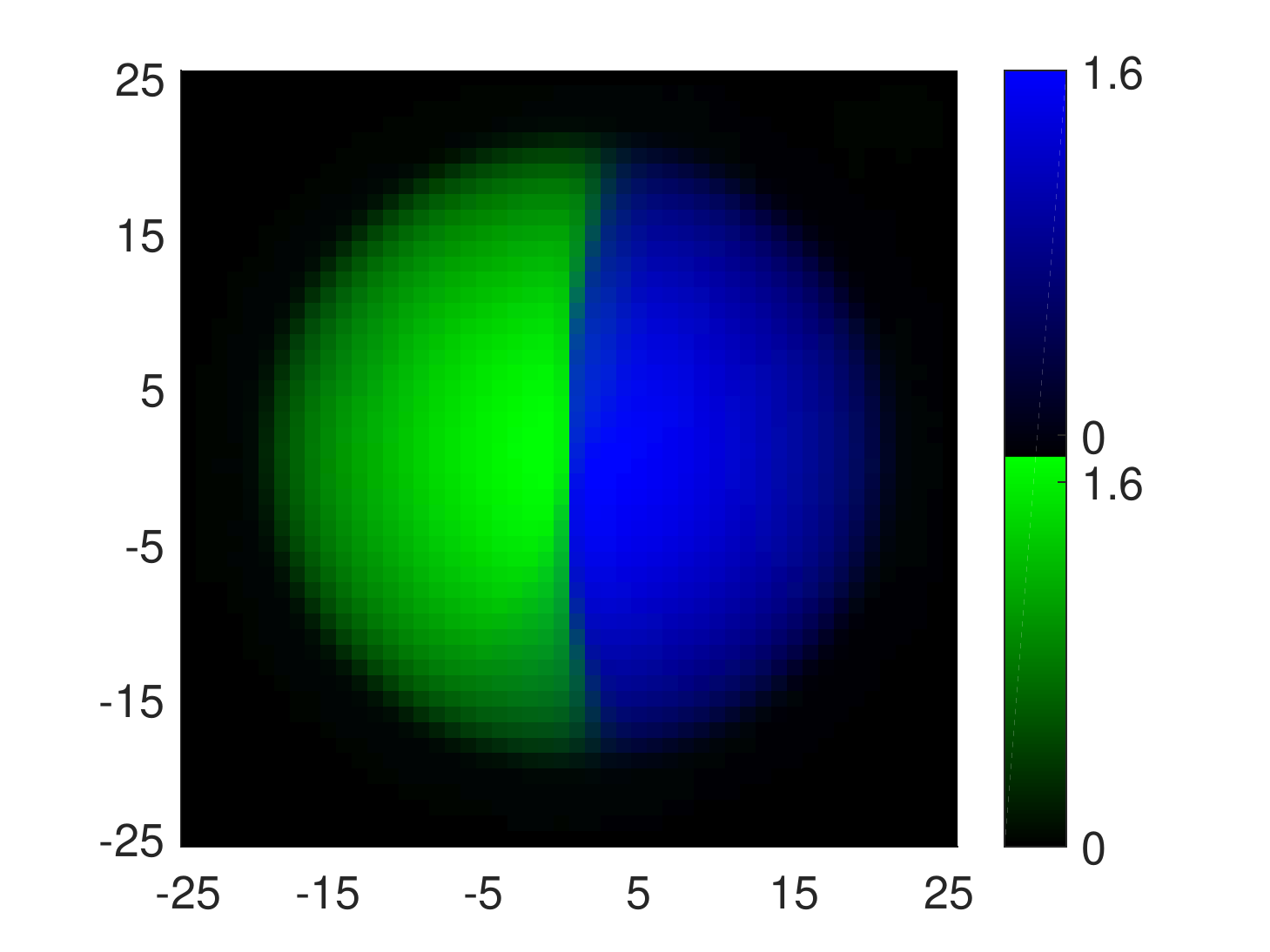}
\end{minipage}
\begin{minipage}[t]{.49\textwidth}
\includegraphics[width=\textwidth]{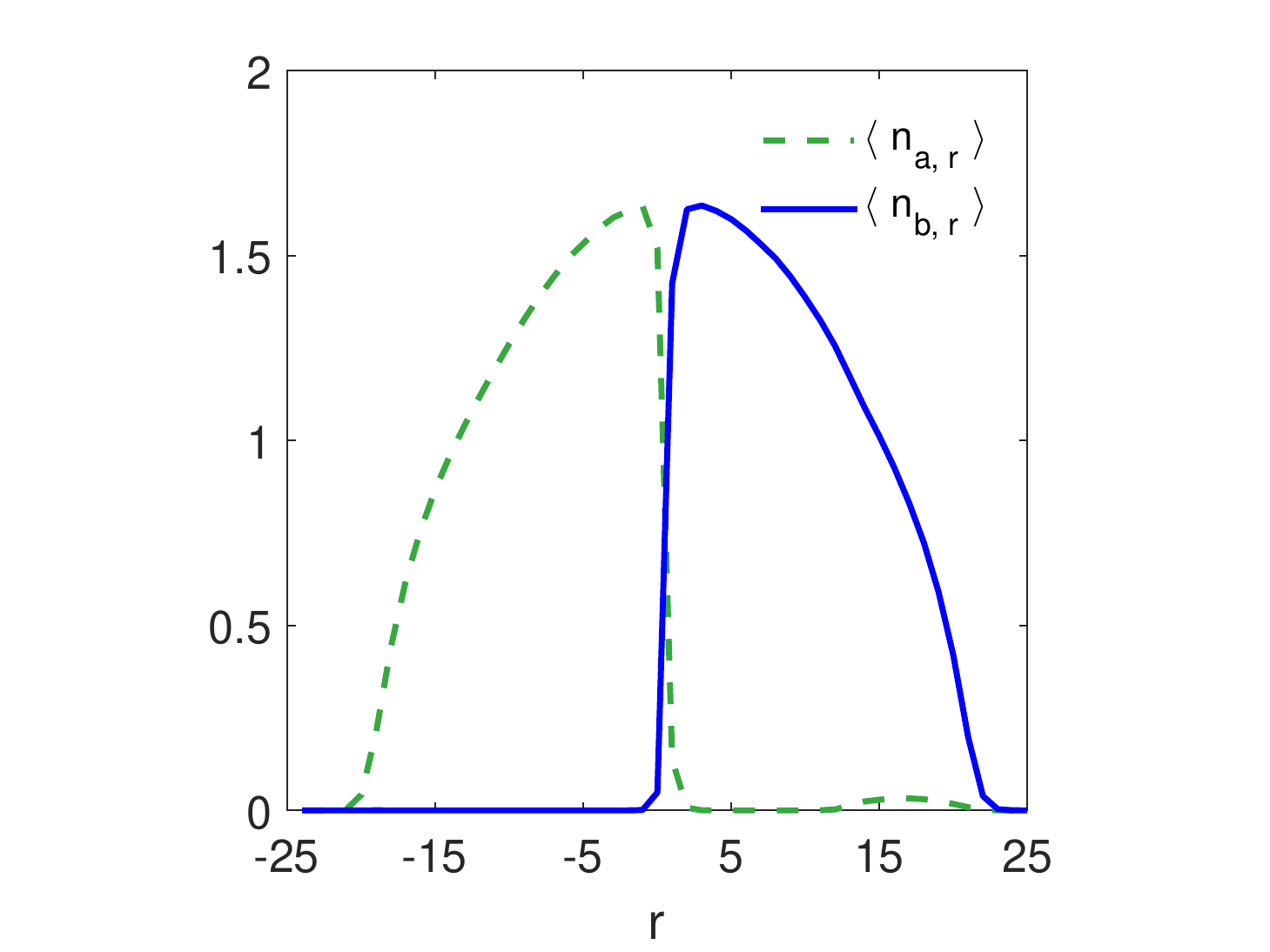}
\end{minipage}
\caption{dSF phase ($U/t=10$, $U_{ab}/t=15$, $N_a=639$, $N_b=641$) confined in a Harmonic trap ($\omega_H/t=0.12$). Left-Panel shows density maps for both species over the entire trap. Right-Panel shows a radial section of density maps, perpendicular to the boundary.}\label{figsp1}
\end{figure}

Here we present further details in our analysis on quantum demixing effect in mixtures of two bosonic species. In particular, we wish to provide further information for the case of non-commensurate filling, analyzing the system in non-homogeneous conditions in the presence of a harmonic trap. In the following we consider balanced populations.
In the presence of a harmonic potential, the chemical potential of species $c=a,b$, occurring in the free energy $F=H- \sum_i \sum_c \mu_{ci}n_{ci}$, transforms according to
\begin{equation}
\mu_{ci}=\mu_{c} - \omega_H {{\vec r}_{i}}^{\,\,2}
\end{equation}
where $\omega_H$ is the strength of the harmonic trap (expressed in unit of the hopping amplitude $t$), and ${\vec r}_i$ the position vector of lattice site $i$.
This leads to a variable local filling factor $n_{{\vec r}}$ through the lattice, allowing us to explore in a more complete way the influence of the filling variation on the phase diagram of the system.
As we showed in the main paper, demixing persists away from filling $n = 1$, as long as $U_{ab}>U$. This is even more evident in the trapped case.  In figure \ref{figsp1} we show the density map of a demixed superfluid in which a harmonic confinement forces the mixture to occupy only the central regions of the lattice with radially decreasing filling. In the right panel of figure \ref{figsp1} we display the demixed density profiles of the two species for a radial section perpendicular to the interface between the two superfluids. It is evident that demixing persists at any value of the radial coordinate and for different values of filling.
It is worth noticing that, as predicted in \cite{segregPhase1998} for a mixture of two Bose-Einstein condensates in a harmonic trap, the overlap energy is minimized when the boundary is a straight line.

\begin{figure}[h!]
\includegraphics[width=0.8\textwidth]{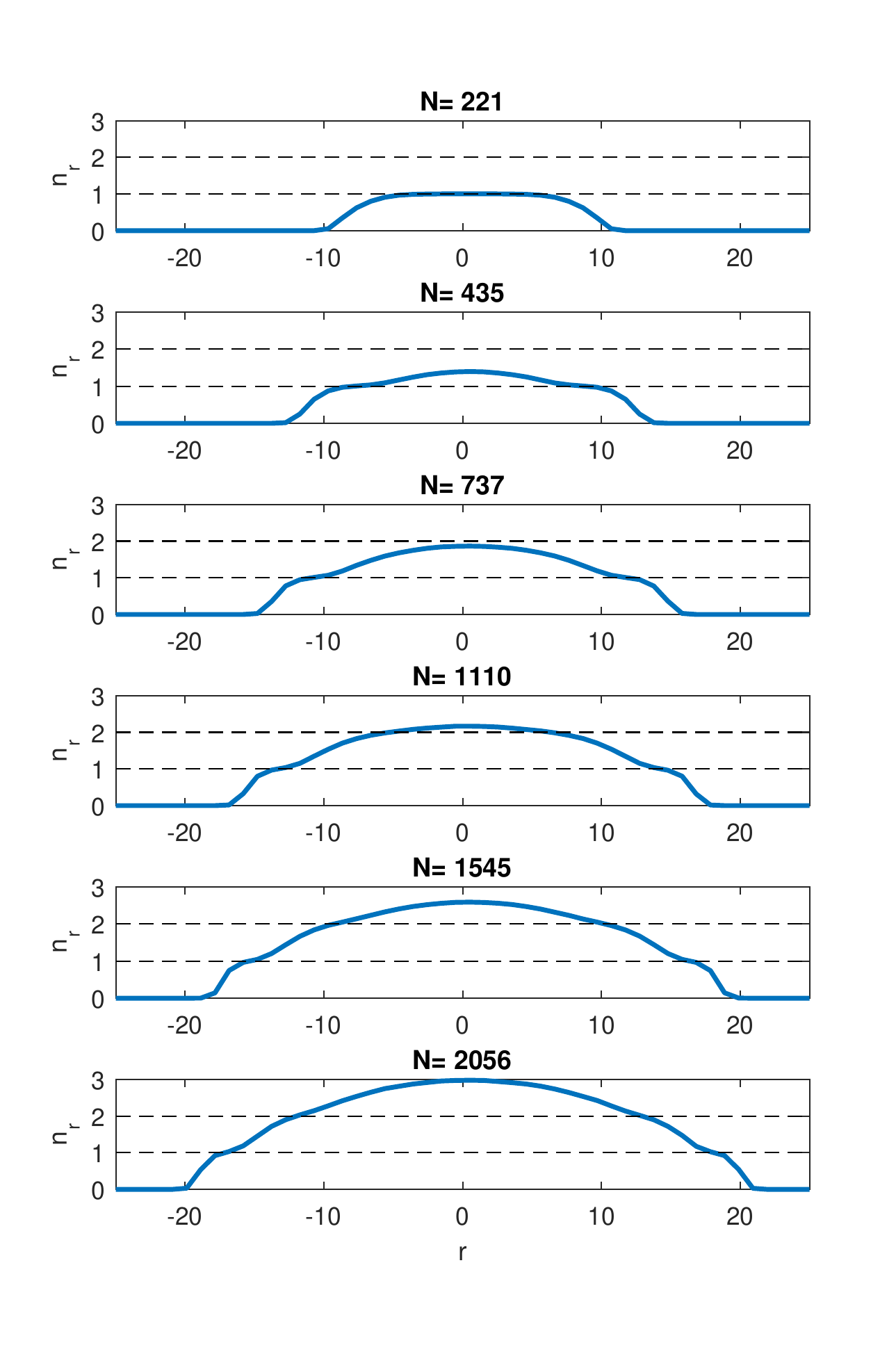}
\caption{Radial profiles of total density $n_{{r}}$ for increasing number of particles at $U/t=20$, $U_{ab}/t=17$ and $\omega_H/t=0.12$. }\label{figsp2}
\end{figure}

Furthermore, for higher values of the inter- and intra-species potentials, within the SCF region of the phase diagram (Fig. 1 in the main text), we observe the presence of the well-known spatial shell structure \cite{shell1,shell2,shell3}. For increasing total number of particles in the system, we observe a spatial structure in which shells of 2SF phase intercalate SCF-phase shells, depending on the value of the local filling factor. The different panels of Fig. \ref{figsp2} show density profiles for different values of $N = N_a + N_b$ but same parameters $U/t=20$, $U_{ab}/t=17$ and $\omega_H/t=0.12$. From top to bottom, radial profiles of total density with increasing boson number are displayed. By increasing $N$, the spatial shell structure appears, revealing the alternate shells of the two phases, with the SCF characterized by a density plateau. We do not observe a plateau at $n=2$ in agreement with reference \cite{mishra} where the formation of SCF phase at double-commensurate filling is observed for larger values of $U$.
